\documentclass{article}
\usepackage[left = 1.5in, right=1.5in, top=1.0in, bottom=0.5in]{geometry}
\usepackage{graphicx}
\usepackage{amsmath}
\usepackage{setspace}
\usepackage{fancyhdr}
\title{Developing a computational model of blood platelets with fluid dynamics applications}
\author{Vijay Viswanathan}
\date{March 2012}
\begin{document}
\pagestyle{fancy}
\fancyhead{}
\renewcommand{\headrulewidth}{0pt}
\newcommand{\comment}[1]{}



\begin{titlepage}
\maketitle
\begin{center}
\verb=vijayv@andrew.cmu.edu=.

\end{center}
\end{titlepage}

\newpage
\lhead{Table of Contents}
\tableofcontents
\newpage
\lhead{Vijay Viswanathan}
\setcounter{secnumdepth}{1}


\section{Introduction}
\emph{Note: this paper was originally submitted to the Siemens Competition for Math, Science, and Technology, so all conventions follow Siemens guidelines.}

This paper considers studying blood platelets using mathematical analysis and methods. Platelets, also known as thrombocytes, play a key role in blood clotting. The role of these entities in strokes, myocardial infarctions, and coronary artery disease add to the importance of blood platelets. The chemical explanations of blood platelet activation and coagulation are rather well understood. Still, the mechanics behind platelet aggregation is largely unknown. According to researcher Constantine Pozrikidis of the University of Massachusetts at Amherst, ``While the biochemical origin of the adhesion kinetics has been well characterized, the non-spherical platelet shape has discouraged the mathematical modeling of the adhesion process by elementary methods of particulate hydrodynamics.'' \cite{pozrikidis} The fact is that the shape of platelets is derailing the hydrodynamic simulation efforts of blood platelet activation, aggregation, and adhesion, and this is hampering scientific progress in hematology.

\begin{figure}
 \centering
\includegraphics{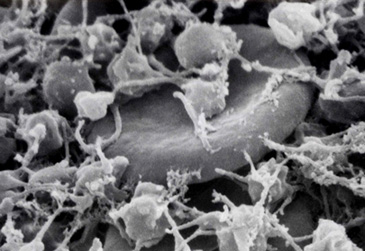}
 \caption{As a graphical aid, here is an image \cite{okla}, not taken by the author, which demonstrates platelet aggregation around a red blood cell.}
\end{figure}

Due to the microscopic nature of blood platelets (the mean platelet volume for an average 65-year old is only 10 femtoliters \cite{mpv} - 100 millionths of a cubic millimeter) accurate laboratory results involving thrombocytes are very difficult to achieve. A newly developing method to approximate these results is computational fluid dynamics (CFD). Using methods of CFD, complex experiments involving fluid flows, turbulence, and interactions between liquids, gases, and solids can be solved using numerical algorithms. However, as Pozrikidis states, mathematical and hydrodynamic models of platelets are generally undeveloped, and this hinders the growth of CFD methods with regards to thrombocytes.

When blood platelets enter the coagulation cascade and aggregate with other platelets, they undergo the process of platelet activation by thrombin. Morphologically, this makes the shape of the platelet shift from an ellipsoid-like object \cite{vwf} to a convex, irregular body. While unactivated platelets are easily representable and have been modeled in the past, activated platelets are more challenging to computationally work with. The reasons for this are twofold: activated platelets have a highly irregular shape, and due to the nature of the platelet aggregation process, very few mechanical results of individual platelets have been found.

In developing a CFD simulation, either the geometry of the solids and fluids in question or more intrinsic characteristics (e.g. viscosity of fluids, density of solids, etc.) must be known. Owing to the complex shape of platelets, the former path becomes computationally inefficient. Similarly, very few existing CFD simulators offer support for shapes which lack a simple functional representation.

 Regarding the intrinsic characteristics of the objects in a simulation of platelets, one primary unknown variable is the drag force caused by the platelet. Again, the blood platelet's odd shape makes evaluating the drag coefficient ($c_d$) rather difficult from a theoretical standpoint. 

In response to the gaps present in both pathways towards a CFD simulation involving thrombocytes, I decided to answer the question of how one can develop a computational model of blood platelets for fluid dynamics purposes. 

This process took three stages. 
The first step in the process of developing the model was outlining the mathematics behind the platelet geometry, both in two-dimensions and in three-dimensions. Next, I developed a method to computationally ``match'' any platelet image to the analytical expressions that represent the shape of the platelet. Finally, the platelet images were utilized to attain the drag coefficient for each blood platelet analyzed. This could become extremely useful if scientists hope to design personalized devices (like stents) or  if pharmacists hope to create computationally designed antiplatelet drugs for patients based on their specific platelet characteristics. This is currently unachievable, but up-and-coming simulations using the knowledge enumerated in this paper could help this dream.
Hopefully accurate CFD simulations could feasibly replace arduous laboratory experimentation. 

\section{ Methods} 

\emph{Note: all programming was performed in C++, primarily using default libraries in Visual C++ 2010 Express, as well as the CMUGraphics Library} \cite{cmu scs}. \emph{Any functions which do not belong to default libraries or CMU graphics were written by the author of this paper. Similarly, all two-dimensional graphing was performed on author-generated software while all three-dimensional graphing was done on Mathematica.} 

The methodology of this project lies in three phases:

\textbf{Phase 1: Two-dimensional modeling.} During the process of achieving a workable three-dimensional model, the first step (for simplicity) was modeling in two-dimensions. Initially, I experimented with the feasibility of taking a platelet image and interpolating the points on the edges of the image with low-order polynomials. However, it became clear that while these formulations were rather faithful to the initial platelet image they were both unnecessary and inefficient. For one, there existed some portions of the shapes (such as vertical lines) which could not be modeled with polynomials with the same variables. Additionally, such complex curve-fitting could not be generalized into an effective framework with a set number of parameters, and the time effectiveness of precise interpolation in three dimensions would seriously stifle the relevance of such an algorithm. 

\begin{figure}
 \centering
\includegraphics{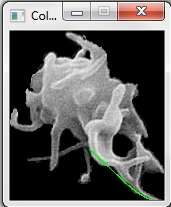}
 \caption{Here is the first approach of the two-dimensional fitting phase, where for every section of the edge of the image, polynomial interpolation created a polynomial approximation.}
\end{figure}

Next, I developed a general framework using the classic curve in polar coordinates known as the ``polar rose'' but generalized it to fit the typical model of a blood platelet. A platelet in a simplified two-dimensional visual looks like an invisible circle with projections of different lengths and widths emerging radially from it. To make the polar rose have this inside circle, a ``generalized polar rose'' which randomly varies the parameters of individual rose petals was created. Furthermore, due to the unavailability of existing software capable of graphing piecewise, polar functions, I designed and implemented my own graphing utility capable of randomizing parameters and graphing piecewise-defined polar roses.

\begin{figure}
 \centering
\includegraphics[scale=0.4]{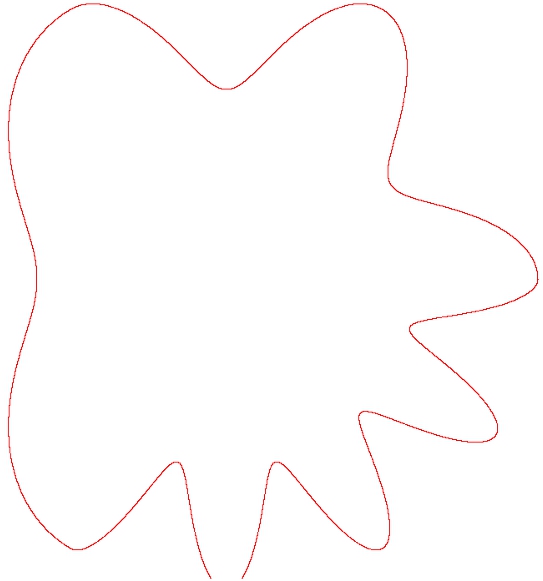}
 \caption{Above  is the second approach in the two-dimensional phase, where using the generalized polar rose model, a platelet-like object was generated }
\end{figure}

\textbf{Phase 2: Three-dimensional expressions.} The initial stage of this phase was to implement the equation of an ellipsoid. This is the core of the platelet model, and projections emerge outwards from the surface of this ellipsoid.

The second stage for the three-dimensional mathematical framework of the platelet morphology was developing the equations of the projections.
To improve smoothness between these surface projections and the underlying discoid structure, a logarithmic function is used for the surface projection. By solving a simple differential equation, the function representing the projections can be made to fit with the ellipse in a manner which maintains differentiability for the most part. In order to generate a three-dimensional surface from this 2D curve, this function was revolved around the z-axis and surfaces of revolution were employed. Additionally, parameters that allow the changing of the length and width of each projection were exercised.

\begin{figure}
  \centering
\includegraphics[scale=0.6]{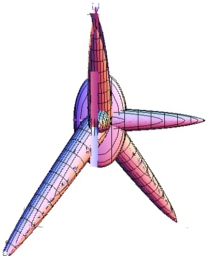}
  \caption{Pictured is a sample model of a platelet using the framework developed in this paper.}
\end{figure}

In order to allow these projections to emerge in all directions rather than exclusively upwards, different approaches were taken. Initially, spherical coordinates were used as a simply mode of rotating the upwards projection to many different direction. However, after converting the $z(x,y)$ function to an $r(\theta,\phi)$ function in spherical coordinates, the function could no longer be solved explicitly (i.e. it was not possible to isolate the $r$ variable from the other two variables). By consequence, while rotations would have been easy to perform, numerical algorithms would be required to evaluate the function in spherical coordinates. Then, cylindrical coordinates were engaged to avoid the complications found in spherical coordinates. However, once the explicit form was achieved in cylindrical coordinates, rotations could only be done along one angle, and this made cylindrical coordinates impractical. To maintain simple and explicit functions, ultimately rotational matrices had to be applied. This rotational matrix was a reduced version of the traditional three-angle rotation matrix in three dimensions, because one of the Euler angles (the ``roll'' angle) was not relevant to the rotation of projections.

\textbf{Phase 3: Computer program and drag coefficient calculation}

\emph{Note: Specific figures and screenshots referred to are in the Illustration section.}

This portion of the work required writing a computer program in C++ which facilitated the process of matching any platelet to a mathematical representation. The platelets were inputted with six different, orthogonal views. For example, an object in three-dimensional Cartesian space with variables (x,y,z) could be understood by viewing the images through the (x,y) plane, through the (x,z) plane, through the (y,z) plane, and through the view from the opposite face of each plane. Additionally, due to the difficulty and inaccuracy of completely automating the process, data by the user was manipulated into a model.

Furthermore, the specific dimensions of the ellipsoid resting within the projections had to be found.

\begin{figure}
  \centering
\includegraphics[scale=0.9]{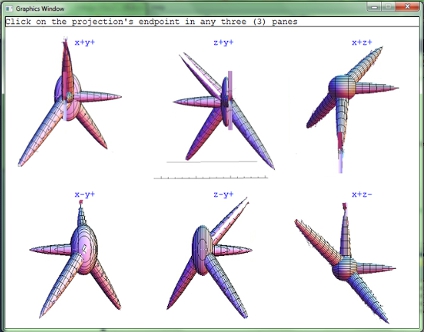}
  \caption{Pictured is the title page of the C++ program. This allows the user to specify which projections he or she wants to use - consequently, the data is analyzed to return a working model for that projection.}
\end{figure}

The last endeavor of this phase was to find the drag force created against fluid flow by blood platelets. In this case, there was no existing literature describing the drag force exerted by an isolated platelet in a flow, thus necessitating the use of a theoretical basis for drag coefficient calculations. Tran-Cong, Gay, and Michaelides \cite{pudn} in 2003 created a method to approximate the drag coefficient ($c_d$) for any object in a fluid flow (with the condition that the flow's Reynolds number had to be between 0.15 and 1500.

Most of the variables in the $c_d$ formula were applicable to all platelets and could be found in recent literature.
My written code used algorithms to determine two of the values required in this theoretical $c_d$ formula.

\section{Results} 

\textbf{Results from Phase 1: }   Fig. 2 \comment{GET THIS IMAGE ON HERE} displays a picture of a platelet image with the polynomial interpolations graphed over it. This shows the primary result for this stage, when the goal was to curve-fit every section of the image's border to a polynomial. 

The next result gained was the two-dimensional model using the idea of a generalized polar rose. The equation for the generalized polar rose is $r=a \cdot cos(b \theta + \phi) + c$, where $a$ indicates projection length, $b$ indicates projection width, $\phi$ rotates the curve, and $\theta$ is the variable for the function. The written program randomized the lengths and widths while connecting the different projections in a smooth manner. This was done by first splitting the total, $360^{\circ}$ curve into subsections of different widths (done by reducing the $b$ value in the generalized polar rose equation so that the quantity of petals in the defined interval is a positive integer). Then, within each section of identical width, individual petals were given random heights. This simple method created a random platelet each time the program was run.  Figs. 3 and 4 display sample random drawings made by the program to approximate the two-dimensional appearance of a platelet. In order to display these piecewise polar functions, a simple polar grapher with support for separate pieces on different intervals was developed. In Figs. 3 and 4, the randomized equations are shown as graphed on this piecewise, polar-curve graphing utility.  

\begin{figure}
 \centering
\includegraphics[scale=0.5]{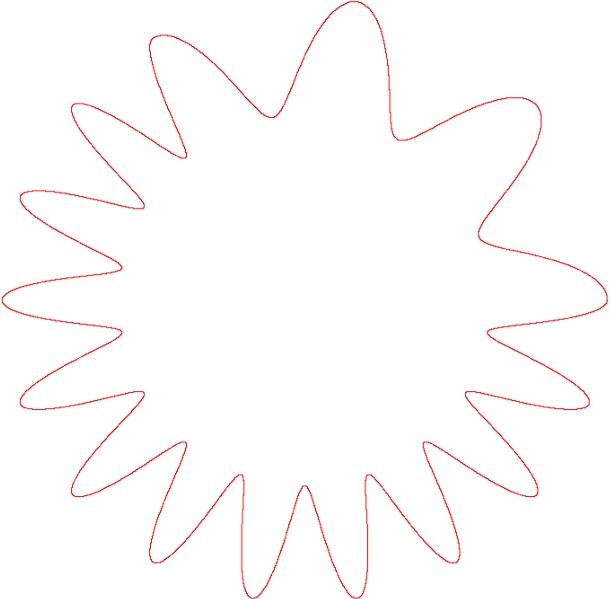}
 \caption{Attached is another generalized, randomized polar rose, of the equation $r=a \cdot cos(b \theta + \phi) + c$. }
\end{figure}

\comment{//////////////////////////////////////////////////////////////////////////////////////////////////////////////////////////////////////////////////}

\textbf{Results from Phase 2: }
The initial stage of this phase was to implement the equation of an ellipsoid. The simple, Cartesian equation for an ellipse with semidiameters $a$, $b$, and $c$ is $\dfrac{x^2}{a^2}+\dfrac{y^2}{b^2}+\dfrac{z^2}{c^2}=1$. This is the core of the platelet model, and projections emerge outwards from the surface of this ellipsoid.

In developing the three-dimensional projection functions, a two-dimensional basis function had to be made. The chosen function type was the logarithmic function. This decision was determined after solving the differential equation $f'(0)=1/0$ (made so that when rotated the point at 0 wouldn't be a cusp, but rather a smooth end). One possibility for $f'(x)$ was the function $f'(x)=1/x$. Integrating on both sides, the resulting function is $\int f'(x)\hspace{2 pt} dx=\int 1/x \hspace{4 pt} dx \Rightarrow f(x)=log(x)+C$. In order to generate a three-dimensional surface from this 2D curve, this function was revolved around the z-axis. As a method finding the equation for the surface of revolution, the radius of the circular level sets of the surface was set to vary with y in a logarithmic manner.  This level set method served as the process to find the projection's surface equation. Furthermore, to adjust the length (denoted by \emph{l}) and width (denoted by \emph{w}) of the projections and for the projections to be outwardly projecting, the specific two-variable function chosen was $y^2+z^2={\dfrac{(log[-x+l])^2}{w}}$. The projection surface that resulted can be represented by the original function $z=\pm \sqrt{\dfrac{(log[-x+l])^2}{w}-y^2}$. 

For these projections to be fitted to face any direction, rotational matrices had to be applied. To determine the general form for rotating the function by two angles, $\theta$ and $\phi$ (colloquially known as pitch and yaw in terms of Euler angles) the rotation matrix must be multiplied with the vector of the original parametric function. The matrix shown below is the rotation matrix when $\psi$, the third Euler angle representing the ``roll'' motion, was set to 0.
\[
 \begin{bmatrix}
  cos \theta cos \psi & sin \psi & sin \theta cos \psi \\
  cos \theta sin \psi & cos \psi & sin \theta sin \psi \\
  -sin \theta & 0 & cos \theta
 \end{bmatrix}
\]

By consequence, the matrix multiplication between the rotation matrix and the vector representing the function is:
\[
 \begin{bmatrix}
  cos \theta & sin \phi sin \theta & cos \phi sin \theta \\
  0 & cos \phi & -sin \phi \\
  -sin \theta & -sin \phi cos \theta & cos \phi cos \theta
 \end{bmatrix}
 \begin{bmatrix}
  x \\
  y \\
  \sqrt{\dfrac{(log[-x+l])^2}{w}-y^2} \\
 \end{bmatrix}
=
\]
$z(x,\theta,\phi, l, m)$=\[
\begin{bmatrix}
  u cos \theta + v sin \phi sin \theta + cos \phi sin \theta \\
  v cos \phi - sin \phi \sqrt{{\dfrac{(log[-x+l])^2}{w}}-y^2} \\
  -x sin \theta - y sin \phi cos \theta + cos \phi cos \theta \sqrt{{\dfrac{(log[-x+l])^2}{w}}-y^2} \\
 \end{bmatrix}
\] 

$z(x,\theta,\phi, l, m)$ represents a projection with length $l$, width $m$, and rotated angles $\theta$ and $\phi$ from the z-axis.

\begin{figure}
  \centering
\includegraphics[scale=0.5]{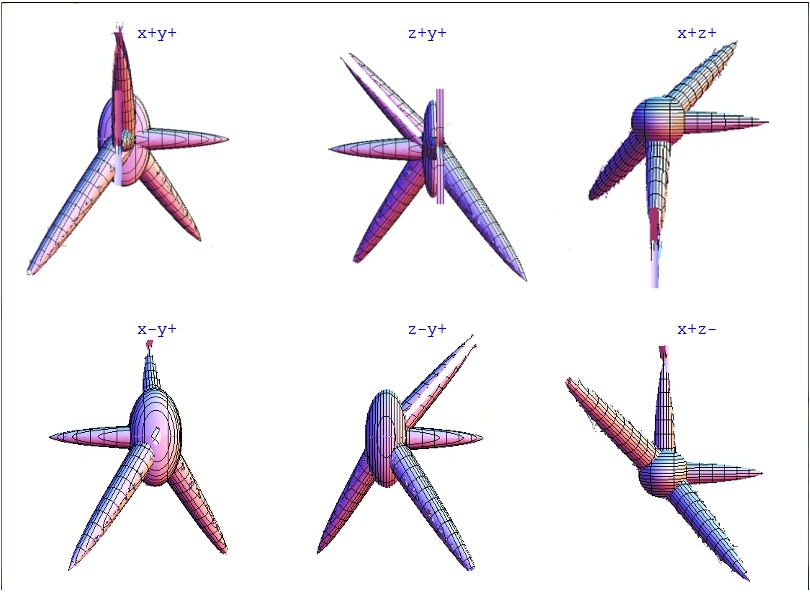}
  \caption{Here are all 6 views of the platelet in Figure 5.}
\end{figure}

\textbf{Results from Phase 3: }
\emph{Note: Specific figures and screenshots referred to are in the Illustration section.}

\begin{figure}
 \centering
\includegraphics[scale=0.5]{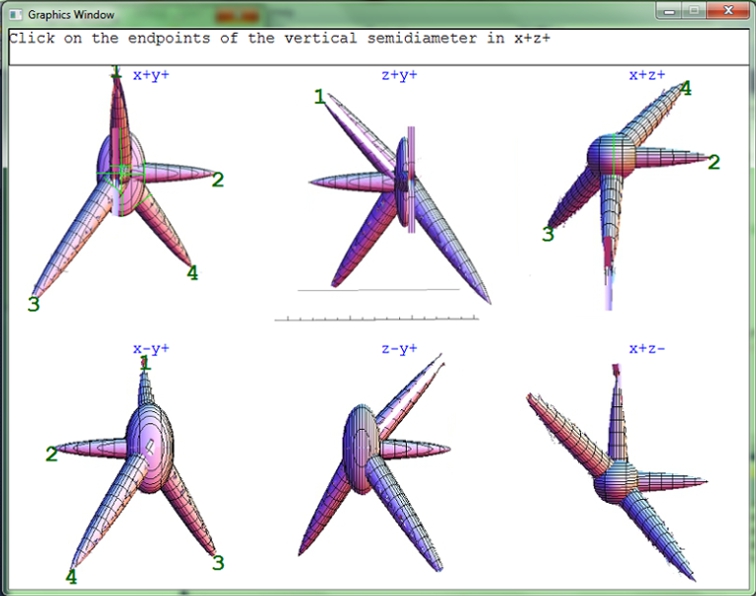}
 \caption{Here is one of the later stages in the program. The thin green lines show the ellipsoid dimensions as well as the width of the projections.}
\end{figure}

The next step was developing the computational aspects of the program. Specifically, the program was supposed to fit the developed mathematical model and match the parameters to any inputted platelet image (represented with six different two-dimensional views of the same three-dimensional object). The program had to gather data on the length and width of each platelet, the dimensions of the underlying ellipsoid beneath the projections, the area of the platelet from one view, and the perimeter of the platelet from the same view. 

The application of the six-view method for analyzing the platelet image became difficult as these specific views have never been taken before. As a result, I approximated the shape of a platelet by taking the existing mathematical model and randomizing the parameters. This created a varied shape which was a suitable replacement for the blood platelet morphology.

The first action was developing an interface on which a user could aid the computer in the detection of the projections. Automatic detection of these projections took too much runtime and was not sufficiently accurate given the typical low-contrast of the images, so significant user input was required. Initially, the six sides of a blood platelet were juxtaposed on the same graphics window. Then, the user clicked on the point representing the endpoint of the same projection in three different views - this way $x$, $y$, and $z$ coordinates could be shown for each endpoint. In the same function, a utility for determining the width of each platelet was defined by allowing the user to click on the endpoints of each projection's base.

\begin{figure}
\centering
\includegraphics[scale=0.8]{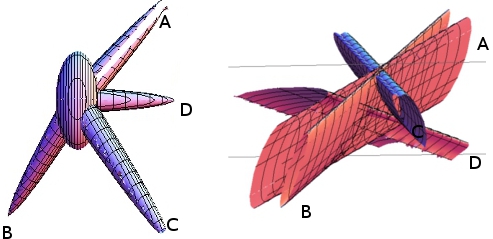}
  \caption{Comparison between the initial platelet image and what was later detected. Issues with a third-party software account for the apparent difference, but fundamentally both images have the same elements. Also, the projections are not correctly bounded in the second image, as based on the mathematics of the model. That is, projection A should be only show the right half side, projection should only show its left half side, etc.}
\end{figure}

Moreover, the dimensions of the underlying ellipsoid had to be discovered. Again, manual detection with automatic computation was favored, and the user clicked on the endpoints of each of the three semidiameters of the ellipsoid. As a result, the $a$, $b$, and $c$ components in the ellipsoid equation, $\dfrac{x^2}{a^2}+\dfrac{y^2}{b^2}+\dfrac{z^2}{c^2}=1$, were identified. In order to compare these equations with real-life data, it was found that the length of 1 pixel in the graphics window corresponded to 10.62 nanometers, and this leads to accurate predictions of platelet size.

The final method in this work was the determination of the drag coefficient for any inputted blood platelet. In 2003, Tran-Cong, Gay, and Michaelides \cite{pudn} developed a formula describing the drag coefficient of an irregular object. The authors also mentioned that this formula is consistent with experimental results for $0.15 < Re < 1500$. Their formula is shown below in Equation 1.

\begin{equation} c_d=\dfrac{24}{Re} \dfrac{d_A}{d_n}[1+\dfrac{0.15}{\sqrt{c}}(\dfrac{d_A}{d_n} Re)^{0.687}]+ \dfrac{0.42 (\dfrac{d_A}{d_n})^2}{\sqrt{c}[1+4.25 \times 10^4 (\dfrac{d_A}{d_n} Re)^{-1.16}]} \end{equation}

In Equation 1, $Re$ is the Reynolds number of the flow, and for blood flow in a human artery, the average Reynolds number is 250. $d_n$ is the nominal diameter of the object, and $d_n = (\sqrt{\dfrac{6V}{\pi}})^3$, where V is the particle volume The specific value for V that we use is 10.0 femtoliters \cite{mpv}. $d_a$ is the volume-equivalent-sphere-diameter.  $d_a = \sqrt{\dfrac{4A_p}{\pi}}$ (where $A_p$ is the projected area of the object in the direction of motion). Lastly, $c$, the particle circularity, can be found with the equation $c =\dfrac{\pi d_a}{P_p}$, where $P_p$ is the projected perimeter of the object in the direction of motion.

The program that was developed found both $A_p$ and $P_p$, and the remaining parameters either depended on $A_p$ and $P_p$ or were found in other publications. The drag coefficient calculated for the platelet shown in Fig. 4 and 5 \comment{WHAT IS FIGURE N? INSERT IT} was 2.874. This corresponds with drag coefficients for more hydrodynamic shapes, thus supporting the claim.

Finally, to make these results applicable to real medical technology, the scale between the length of one pixel in the graphics component and physical length of platelets had to be found. Using the $A_p$ variable, $sqrt{A_p}^3$ roughly represents the volume of an object with all sides of area $A_p$. Comparing $\sqrt{A_p}^3$ with the mean platelet volume (assumed to be 10.0 femtoliters), are roughly 0.0000043 femtoliters per voxel. These means$ 4.28 \times 10^{-21}$ liters per voxel. By taking the cubic root of both $ 4.28 \times 10^{-21}$ and the liters per voxel unit, the result is that the length of 1 pixel roughly corresponds to $1.62 \times 10^{-8}$. By this measure, the average length of 1 pixel of the platelet is roughly 16 nanometers.

\section{Analysis} 
\comment{            Be thorough, the discussion is the essence of your paper. Tell your readers exactly what
                            you did and thought. Compare your results with theories, published data, commonly held beliefs,
                            and expected results. Discuss possible errors. How did the data vary between repeated observations
                            of similar events? How were results affected by uncontrolled events? What would you do differently
                            if you repeated this project? What other experiments should be conducted?                 }
Going into this project, I had expected to design a basic mathematical model which would then go into a coarse-grained molecular dynamics model of platelet activation. However, my model was not discrete enough for a dissipative particle dynamics approach to coarse-grained molecular dynamics (MD). Moreover, were my project to become a component of an actual MD simulation, the runtime of the simulation would literally take years, even if the processes were parallelized on high-performance computing clusters. 

The transition from a computational model with MD applications to one with primarily CFD results was rather smooth given the nature of my model. The equations that I provided were continuous and, for the most part, smooth wherever defined. While my use of the logarithmic function in a model that hoped to be differentiable could be somewhat alarming, $log(x)$ is differentiable on all orders for all values on which $log(x)$ is defined. By consequence, the surface of the geometric platelet model is almost completely differentiable.

My findings could not be referenced with external data because of a lack of published material  on the fluid mechanics of thrombocytes. While some sources do reference the drag force of platelets in their calculations\cite{rube}, never do they spefically mention values or equations for calculating the drag forces and coefficients. Similarly, Karniadakis et. al.\cite{karn} mention drag coefficients of platelets as generally varying inversely with the Reynolds number, but there is an absence clear documentation of $c_d$ values. This chasm in knowledge is clear when looking at some influential blood platelet simulations. For example, Eckstein and Belcagem \cite{miami} describe a model of platelet transport with drift and diffusion, but they also write, ``No extra drag or interaction with the wall (beyond that included as a part of the drift) was included,'' with respect to platelets in their simulation. 

Rather than being supported by these other findings, my work enhances these findings. For example, in the poster by Karniadakis et. al. \cite{karn}, more rigorous calculations of the drag coefficient could be computed without significant experimental issues. Similarly, in Eckstein and Belcagem's paper, they could improve the scope of their work by including not only drift and diffusion in the model, but also drag. Many avenues of research can be expanded upon with this computational method of discovering drag, because this reduces and in the case of blood platelets eliminates the need for many experimental procedures relating to these calculations.

The expected result was for the mathematical model be reasonably faithful in its fit to a real platelet while not impeding on the model's efficiency. However, due to the lack of sufficient platelet images, this could not be tested - only the solution could be given. Still, when comparing the approximated platelet (generated by randomizing an ellipsoid with projections around it), while initially the result looks much busier than the original input, closer analysis shows that this effect is due to the different scales required for each parameter. The parametric equations have two parameters, $u$, and $v$, but in Mathematica the two parameters are always treated in the same interval for all parametric surfaces in the same graphing window. As a result, this does not account for the different intervals of graphing required for each surface ($u$ and $v$ should only be defined for each projection until the very tip of the projection can be seen - after that surface extraneities emerge).\comment{YOU HAVE TO PUT THIS IN.} 

When varying different approximations of platelets as the input of the program, there was generally strong consistency. Only if the variation between projections  was very strong would real issues arise, and even that was largely due to issues with the Mathematica graphing windows rather than anything else. The equations themselves were correct based on comparison between the resultant equations and the input equations that defined the ``platelet'' which served as the input for my code.

Ultimately, this work is unique and is a contribution to science because it uses easy-to-use computational methods to build on very difficult experimental problems. For example, stents to prevent infarctions could be designed using CFD software such that for a person with a rare blood disorder, the safety and efficiency of the stent could be assured. Similarly, my mathematical model could be used to develop normative criterion for ``safe" platelet structure. Patients not falling completely into these normative intervals could be investigated further and diagnosed. For example, patients whose average projection length exceeds the national average would require further examination.


\section{Conclusion and Future Work} 
My work accomplished the two goals that were set out. I developed analytic expressions representative of blood platelets in varying stages of activation and developed code that fit these expressions to faithfully represent any given blood platelet. As a result of this work, future computational fluid dynamics simulations involving blood platelets in laminar fluid flows as well as simple turbulent flow. Moreover, because this work provides two different bases for improved simulations (both a geometric model and a solution for the drag coefficient), it provides a solution for simulations on two platforms.

My conclusions are supported by the results in the report alone because of the chasm between what results are computationally possible and what results are achievable through actual experimentation. Experimentally, the small size of blood platelets and the rapid changes that occur to thrombocytes hinder efforts to calculate platelet mechanics. As such, specific results, like the average drag coefficient of an activated platelet, are not available to support my current claims. The only method that could somewhat verify my drag calculations is by comparing my results to results for other, distinct shapes. For example, if the drag coefficient for a triangular prism pointing away from the direction of flow was 1.14 \cite{nasa} then the drag coefficient for a blood platelet would have to be substantially greater due to the highly non-aerodynamic or hydrodynamic shape of the platelet.

However, my methods are certainly not infallible. Most of the testing for my program relied on randomly-generated, analytical models. Therefore, my program fitted a model to an existing image of a model. Still, because of the variety of blood platelet shapes, this should not have caused major problems. Furthermore, some of my algorithms (such as my algorithm to find the perimeter of irregularly shaped images) could have been overestimations or undercalculations.

Experiments that would have to be performed to refine the calculations would necessitate the experimental testing of my theoretical results. When molecular biological techniques, microfluidics, and techniques of microscopy become sufficiently refined, accurate fluid dynamic calculations of blood platelets in blood flow will become achievable. The theoretical results in this paper could then be checked with the experimental results and statistical measures could potentially ensure significance. Likewise, in the future other mathematical frameworks to fit platelets and other related fluid dynamics calculations could be performed and referenced with the current theoretical results to check for accuracy.

Additionally, to improve this research, the detection  of the characteristics of each platelet (like the  lengths  and  widths  of projections  and  the  dimensions  of the  underlying  ellipsoid) must  be efficiently automated.

This research aimed to be more valid than any other currently published, computational results. In fact this seems to be the case, as an exhaustive literature review found no clear mathematical model for blood platelet morphology and results for the drag coefficient of blood platelets are nonexistent. 

My future for this work is that simulations can be used to develop the way medicine is approached. Through this paper I attempted to patch up many of the holes that inhibit accurate and efficient computational fluid dynamics simulations. In the future there would be two primary concepts which could be expanded upon. Firstly, more sophisticated computer vision techniques could be applied in the creation of a model for medical images. This would completely automate the process and improve general applicability. Second, similar methods of morphological model matching can be used for other aspects of medicine and biology.

\end{document}